\documentstyle[12pt]{article}
\newcommand{\BEQ}{\begin{equation}}
\newcommand{\EEQ}{\end{equation}}
\newcommand{\BEA}{\begin{eqnarray}}
\newcommand{\EEA}{\end{eqnarray}}
\newcommand{\BAR}{\begin{array}}
\newcommand{\EAR}{\end{array}}
\newcommand{\ra}{\rightarrow}
\newcommand{\Ra}{\Rightarrow}
\newcommand{\vph}{\varphi}
\newcommand{\pih}{\frac{\pi}{2}}
\newcommand{\HH}{\cal H}
\newcommand{\la}{\lambda}
\newcommand{\sig}{\sigma}
\newcommand{\Ga}{\Gamma}
\newcommand{\zeile}[1]{\vskip #1 \baselineskip}
\catcode`\@=11
\def\numberbysection{\@addtoreset{equation}{section}
        \def\theequation{\thesection.\arabic{equation}}}
\numberbysection
\begin{document}
\begin{titlepage}
\null
\begin{center}
\vskip 2mm
{\Large {\bf Phase diagram and two-particle
                          structure of the $Z_3$-chiral Potts model}}
\vskip 1cm
G. von Gehlen$^\star$
\zeile{1}
Physikalisches Institut der Universit\"{a}t Bonn \\
Nussallee 12, D - 5300 Bonn 1, Germany \\
\zeile{2}
\end{center} {\large \bf Abstract :}
We calculate the low-lying part of the spectrum of the
$Z_3$-symmetrical chiral Potts quantum chain in its self-dual and integrable
versions, using numerical diagonalisation of the hamiltonian for
$N \leq 12$ sites and extrapolation $N \ra \infty$. From the sequences of
levels crossing we show that the massive phases have oscillatory correlation
functions. We calculate the wave vector scaling exponent.
In the high-temperature massive phase the pattern of the low-lying
levels can be explained assuming the existence of two particles,
with $Z_3$-charge $Q\!=\!1$ and $Q\!=\!2$, and their
scattering states. In the superintegrable case the $Q\!=\!2$-particle has twice
the mass of the $Q\!=\!1$-particle. Exponential convergence in $N$ is observed
for the single particle gaps, while power
convergence is seen for the scattering levels. In the high temperature limit
of the self-dual model the parity violation in the particle dispersion relation
is equivalent to the presence of a
macroscopic momentum $P_m = \pm \vph/3$, where $\vph$ is the chiral angle. \\
\vspace{5cm} \\ \noindent July 7th, 1992 \hspace{95mm} BONN-HE-92-18 \\  \\
Talk presented at the "International Symposium on Advanced Topics of Quantum
Physics",~~Shanxi University,~~Taiyuan,~~China,~~ June 11~-~16, 1992 \\
\\ $^\star$ e-mail-address : UNP02F@ibm.rhrz.uni-bonn.de
\end{titlepage}

\section{Introduction}
The $Z_3$-symmetrical chiral clock model ("$CC_3$-model") has been introduced
in 1981 in order to describe the
transition between commensurate and incommensurate phases of adsorbed gases
on crystal surfaces \cite{Os81,Hu81,denis}. It is a two-dimensional model
defined by the partition function $Z$ and the action $S$:
\BEA  Z & = & \sum_{(a)} \exp(-S) \nonumber \\
      S & = & -\sum_{x,y} \left( g_x \cos[{\textstyle \frac{2}{3}}
 \pi (a_{x+1,y} -a_{x,y})] +g_y \cos[{\textstyle \frac{2}{3}}\pi (a_{x,y+1}
-a_{x,y}) - \vph ] \right).    \label{cc3} \EEA
The $Z_3$-spin variables at sites $(x,y)$ take the values $a_{x,y}=0,1,2$.
{}~$g_x$ and
$g_y$ are the horizontal and vertical couplings, respectively. For non-zero
chiral angle $\vph$ the system is parity-noninvariant in the $y$-direction.
For $\vph\!=\!0$, (\ref{cc3}) is the usual $Z_3$-clock model.
A standard experimental example for a system exhibiting $Z_3$-symmetry and
which may be described by (\ref{cc3})
is monolayer krypton covering of a graphite surface \cite{bak,ein}.
\par
Various techniques have been used in order to determine the phase
diagram of the model: Mean field, Monte-Carlo, renormalization group,
partial transfer matrix diagonalization, and hamiltonian limit finite-size
calculations \cite{cen}-\cite{sev}. As yet,
no full agreement has been reached between various authors on the details of
the phase diagram.
\par The $CC_3$-model is not self-dual. A self-dual model which should show
commensu\-rate-incommensurate phase transitions was proposed and studied
by Howes, Kada\-noff and Den Nijs in 1983 \cite{how}. Their model is defined
by the following quantum chain hamiltonian (or logarithm of the transfer
matrix):
\BEQ    {\HH} = -\frac{2}{\sqrt{3}} \sum_{j=1}^{N} \left\{
 e^{-i\vph/3} \sig_{j} + e^{+i\vph/3} \sig^{+}_{j} + \lambda \left(
 e^{-i\phi/3} \Ga_{j}\Ga^{+}_{j+1} + e^{+i\phi/3}
 \Ga^{+}_{j}\Ga_{j+1} \right) \right\},               \label{H3}      \EEQ
where $\sig_{j}$ and $\Ga_{j}$ are $3\times 3$-matrices acting at site $j$
(the $Z_3$-analogs of the $Z_2$-Pauli matrices $\sig^z$ and $\sig^x$):
\BEQ   \sig_j = \left( \BAR{ccc} 1&0&0\\0&\omega&0\\0&0&\omega^2
       \EAR \right)_{j},  \hspace{1cm} \Ga_j =
 \left( \BAR{ccc} 0&0&1\\1&0&0\\0&1&0 \EAR \right)_{j}  \label{sG}    \EEQ
with  $\omega =\exp(2\pi i/3)$. ${\cal H}$ contains the three parameters $\la,
{}~\phi$ and $\vph$. $\la$ is the inverse temperature,~~$\phi$ and $\vph$
are two chiral angles. For $\phi=\vph$ the model is self-dual.
\par For $\phi=0$ or $\vph=0$ the hamiltonian $\HH$ follows from the
$CC_3$-action (\ref{cc3}) \mbox{\cite{cen,geh2}}, but for general $\vph$ and
$\phi$ the corresponding
two-dimensional lattice model has complex Boltzmann weights, so that there
is no straightforward statistical mechanics interpretation.
Because of some
symmetries of ${\HH}$, it suffices to consider only $0 \le \phi,~\vph \le
\pi$. The spectrum of $\HH$ decomposes into three charge sectors $Q=0,1,2$
because $\HH$ commutes with the $Z_3$ charge operator $\tilde{Q} =
\prod_{j=1}^N \sig_j$ and we may write the eigenvalues of $\tilde{Q}$
as $\exp(2\pi iQ/3)$. Although for $\phi \neq 0$, parity is violated, $\HH$
is still translational invariant (we use periodic boundary conditions
$\Ga_{N+1}=\Ga_1$) and the eigenstates will be labeled by the momentum
eigenvalues $p$:        \BEQ
p =-{\textstyle [\frac{N}{2}]},\ldots,-1,0,1,\ldots,{\textstyle [\frac{N}{2}]}.
        \label{p}    \EEQ
\par Using perturbation expansions and fermionic techniques, Howes {\em et al.}
concluded that the phase diagram of the self-dual model ($\vph=\phi$) contains
three phases:
an ordered and a disordered phase separated by a $Z_3$-Potts universality class
line for low values of $\vph$, and an incommensurate (IC)-phase for
large $\vph$. However, the most exciting result of their analysis (they used
a 10th-order high-temperature expansion) was that for $\vph=\pi/2$
the lowest gap of $\HH$ is a {\em linear} function of $\la$ (the 2nd
to 10th-order coefficients in $\la$ were all found to vanish), just as it is
the case for the Ising model.
\par More precisely, let us define the gaps of $\HH$ with respect to the
lowest ${\mbox Q\!=\!0,\, p\!=\!0}$-level, which we denote by
$E_0(Q\!=\!0,p\!=\!0)$. Observe that
$E_0(Q\!=\!0,p\!=\!0)$ is not always the ground state of $\HH$, since
the model has an IC-phase region where in general the ground state carries
non-zero momentum $p$.
\par   By $\Delta E_{Q,i}$ we denote the energy difference
\BEQ  \Delta E_{Q,i} = E_i(Q,p=0) - E_0(Q=0,p=0)        \label{gap} \EEQ
where $E_i(Q,p=0)$ is the {\em i}~th level $(i = 0,1,\ldots)$ of the charge
$Q$, zero-momentum sector. Then the Howes {\em et al.} result is   \BEQ
\Delta E_{1,0} = 2(1-\la) \hspace{13mm} {\rm for} \la<1. \label{lin} \EEQ
\par  In 1984, v.Gehlen and Rittenberg \cite{geh} looked into the question,
whether also other gaps of the Howes {\em et al.} model might be linear in
$\la$. They found that indeed this is the case. They also constructed
a whole series of $Z_n$-symmetrical models which have gaps linear in $\la$.
These models are defined by the quantum hamiltonians
\BEQ  H = -\sum_{j=1}^{N} \sum_{k=1}^{n-1} \left\{ \bar{\alpha}_k \, \sig_{j}^k
 + \lambda \, \alpha_k \, \Ga_{j}^k \, \Ga_{j+1}^{n-k} \right\}  \label{H} \EEQ
Here $\sig_j$ and $\Ga_j$ are $n\times n$-matrices acting at site $j$ (the
$Z_n$-generalizations of (\ref{sG})) which satisfy
\BEQ  \sig_j \, \Ga_{j'} = \Ga_{j'} \, \sig_j~ \omega^{\delta_{j,j'}}, ~~~~~~
\sig_j^n = \Ga_j^n = {\bf 1},~~~~~~\omega = \exp(2 \pi i/n).    \label{om} \EEQ
The $Z_2$-version of (\ref{H}) is the standard Ising quantum chain.
\par The linearity of gaps is obtained if in (\ref{H}) we choose
\BEQ \alpha_k = {\bar \alpha}_k = 1 -i\cot(\pi k/n).     \label{sui}  \EEQ
At the same time eq.(\ref{sui}) guarantees \cite{geh} that the models
(\ref{H}) are integrable
in the sense that they satisfy the Dolan-Grady conditions \cite{dol},
which are equivalent \cite{dav} to the
Onsager algebra \cite{ons,rr,ahn}. This kind of integrability is now usually
referred to as "superintegrability" \cite{al2,coy}. It ensures that if
$\alpha_k, {\bar \alpha}_k$ satisfy eq.(\ref{sui}), then
                the $\la$-dependence of all eigenvalues $E(\la)$
of $H$ has the special Ising-like form \cite{al2,dav}:
\BEQ  E(\lambda) = a + b\lambda + \sum_j 4m_j\sqrt{1 +\lambda^2 +
 2\lambda\cos\theta_j}.    \label{Isi}  \EEQ     Here $a, b$ and $\theta_j$ are
real numbers and the $m_j$ take the values $m_j = -s_j, -s_j +1, \ldots,
s_j$ with $s_j$ a finite integer. For details on formula (\ref{Isi}) see
\cite{al2,dav}.
\par In 1989 Au-Yang {\em et al.} \cite{au2,au1} derived the hamiltonian
(\ref{H}) from a two-dimen\-sio\-nal lattice model built out of
$Z_n$-symmetrical (but not generally positive) Boltzmann weights. These
satisfy a new kind of Yang-Baxter equations with the spectral parameters
defined
on a Riemann surface of high genus. In a series of papers, the authors of
\cite{au2}-\cite{das} have been able, using in addition a
functional equation \cite{bbp}, to calculate
the complete spectrum of the $Z_3$-hamiltonian (\ref{H3}) in the
superintegrable
case $\phi=\vph=\pi/2$.
\par From the Yang-Baxter equations it follows that (\ref{H}) is also
integrable
if the $\alpha_k$ and ${\bar \alpha}_k$ satisfy the relations
\BEQ    \alpha_k = e^{i(2k/n-1)\phi}/\sin(\pi k/n),~~~~~~~
  \bar{\alpha}_k = e^{i(2k/n-1)\vph}/\sin(\pi k/n),           \label{alpha}
\EEQ
\BEQ  \cos \vph = \lambda \cos \phi                         \label{int}   \EEQ
Except for the superintegrable case, which
is contained in (\ref{int}) for $\phi=\vph=\pi/2$, these equations force
$H$ to be non-self-dual. Calculating the spectrum for the integrable, but not
superintegrable case (\ref{alpha}),~(\ref{int})
is a very difficult task. Nevertheless some initial steps have
been done in attaccing this problem \cite{roan}.
\par In this talk, we shall give some new results on the $Z_3$-model
(\ref{H3}).
We shall not rely on exact solutions and calculate low-lying energy levels
of $\HH$ by numerical diagonalization (using the Lanczos algorithm) of the
chain hamiltonian for
up to $N=12$ sites. The motivation for such a work is two-fold: \\ 1)~
For the superintegrable case, where exact results are available, the numerical
evaluation of the exact formulae is not easy, and straightforward numerical
diagonalization sometimes gives more transparent results on the behaviour of
whole sets of levels. This may lead to the discovery of regularities
which otherwise could be overlooked. Indeed, a good fraction of the progress
in finding the special properties of (\ref{H}) has been made possible
by the extensive finite-size studies performed by the authors
of \cite{geh,al5}.
\\ 2)~Of course, the main merit of the finite-size method is, that it is not
restricted to the superintegrable case and so one can study how the special
superintegrable spectrum emerges from the $\vph$-dependence of the general
spectrum.
\par  In the next Section we shall show that the massive phases are
oscillatory,
Sec.~3 gives a calculation of the corresponding wave-vector scaling
critical index, which sheds new light on the scaling asymmetry noticed in
\cite{al5,ba1}. Sec.~4 then interprets the low-lying spectrum of the massive
phases in terms of "elementary" particles and their composite states. Sec.~5
summarizes our results.

\section{Oscillatory behaviour of the massive phases}
Due to its complex coefficients,
the hamiltonian (\ref{H3})
allows ground state level crossings (the Perron-Frobenius
theorem does not apply). The different phases can be easily distinguished by
the $Z_3$-charge sectors and momenta of the lowest energy levels of $\HH$.
In Fig.~1 for fixed number of sites $N=10$ we draw a map of the $\la / \vph$-
plane (in order not to get lost on three-dimensional figures, in the following
we shall consider the self-dual case $\phi=\vph$ if not explicitly
stated otherwise). We label regions in this plane by the sectors of the
two lowest levels, e.g. $0_1/2_3$ means that the ground state is in the
$Q\!=\!0,p\!=1$-sector and the first excited state has $Q\!=\!2,p\!=\!3$.
In most cases the lines mark alternately cross-overs in the ground state
and in the first excited state.  \vspace{148mm}   \par
Fig.~1 : Map of the lines where cross-overs in the two lowest levels of the
hamiltonian (\ref{H3}) (self-dual version $\vph=\phi$) for $N=10$ sites
occur. The meaning of the labelling of the regions between the cross-overs
is explained in the text.    \newpage
\vspace*{136mm}  \par
Fig.~2: Schematic phase diagram of the self-dual version of the $Z_3$-chiral
Potts model defined by the hamiltonian eq.(\ref{H3}).  \zeile{1}

Inspection of Fig.~1 shows that there are at least 3 different phases
(Fig.~2 gives a general schematic view of the phase diagram):
\begin{itemize} \item Low-temperature phase (I) at $\la>1$ and small $\phi$:
Here the three lowest states are from three different charge sectors
and belong to the zero momentum sector. As $\la$ increases at fixed $\vph$,
the three
levels form a kind of braid and alternate in forming the ground state.
We mention that the gaps between the braiding levels decrease exponentially
both with $N$ and $\la$, as it is expected for the low-temperature phase where
the ground state is threefold degenerate in the thermodynamic limit $N \ra
\infty$ (the next levels follow at the distance of $\Delta E \approx 3(\la-1)$
as has been shown by Baxter \cite{ba1}).
\par In their low-$\vph$-branches
the cross-over lines follow a simple pattern for $\la \ra \infty$:
all these lines tend to fixed values of $\vph$ according to the rule \cite{GK}
(which we verified numerically):   \newpage
\begin{displaymath}
\begin{array}{lll}
 0_0/1_0$~~~(crossing in the ground state)$ &\Ra & \vph = \phantom{2}\pi/N \\
 0_0/2_0$~~~(crossing in the 1st excited state)$ & \Ra &  \vph = 2\pi/N  \\
 1_0/2_0$~~~(ground state)$ &  \Ra &  \vph = 3\pi/N  \\
 1_0/0_0$~~~(1st excited state)$ &  \Ra &  \vph = 4\pi/N  \\
 2_0/0_0$~~~(ground state)$ &  \Ra &  \vph = 5\pi/N  \\
 2_0/1_0$~~~(1st excited state)$ &  \Ra &  \vph = 6\pi/N  \hspace{5mm} etc.
                     \label{phili}
\end{array}
\end{displaymath}
These rules are teaching us that for $N \ra \infty$ there will be an infinite
supply of cross-over lines, which finally will fill the whole region $\la>1$
down to $\vph=0_{+\epsilon}$. We have also verified that for free
b.c. (and also for the integrable version of the model) the same asymptotic
behaviour of the cross-over lines appears. So these cross-overs are not
just artefacts of the boundary condition. We conclude that this phase
is an oscillatory massive phase. In the next Section we shall give evidence
that for increasing $N$ the cross-over lines converge to the line $\la=1$ and
give the calculation of
the wave vector critical index.  \par
\item High-temperature phase (II). Dual to phase (I) this appears at $\la<1$
and
small $\vph$. Here the ground state always belongs
to the sector $Q\!=\!0,p\!=\!0$ so that there are no ground state
level crossings. However, there is a regular pattern of cross-overs in the
first
excited state, which is always in $Q\!=\!1$, but with increasing $N$ and $\vph$
switches to higher momenta. Perturbation theory in $\la$ teaches us that
at $\la=0$ the first excited state moves from momentum $p_0$ to $p_1$ at the
chiral angle   \BEQ   \vph=\frac{3\pi}{N}(p_0 +p_1). \label{mac} \EEQ
Again, this rule ensures that for $N \ra \infty$ there is an infinite supply
of cross-over lines, this time in the first excited level. This gives rise
to a modulation of $\Delta p \neq 0$-pair correlation functions (the
$\Delta p=0$-pair correlations are non-oscillatory).
\item Massless incommensurate phases (III,~IV): In the middle upper part of the
phase diagram there are plenty of ground- and first-excited state level
crossings. Both the charge sectors and the momenta are changing simultaneously.
With increasing $N$ the pattern of cross-over lines becomes more dense. The
loss of translational invariance of the ground state together with a $\la$-
dependent variation of the ground state momentum is sign of the presence of
an incommensurate (IC)-phase. \par Here the information from the exact solution
of Albertini {\em et al.} \cite{al2} on the line $\vph=\pi/2$ provides more
detailed
information than we are able to obtain from our chains of $N \le 12$ sites:
In \cite{al2} it is shown that on the superintegrable line the ground state
looses translational invariance in the interval $0.901293\le \la \le 1/0.901293
$. In \cite{acoy} in addition it is shown that actually for $\la<1$ and $\la>1$
there are two different IC-phases which do not join
smoothly at $\la=1$. \par The appearance of $p\!\neq\!0$ in the
first excited state in the high-temperature region is an indication that
eventually the ground state may acquire momentum too, but $N\le 12$ sites are
not enough to prove that this happens already on the line $\vph=\pi/2$.
\par Interesting information is obtained from our calculations on the
different behaviour
of the self-dual ($\vph=\phi$) and integrable
($\cos\phi=\la\cos\vph$) hamiltonians
for $\la>1$ and $\vph \ra \pi$: In the integrable case we
find that the ground state cross-over lines for $\vph \ra \pi$ concentrate
below $\la \approx 1.6$ and no cross-overs are left at $\vph=\pi$ above
this value of $\la$.
This is in contrast to the self-dual case, where the cross-over lines for
$\vph \ra \pi$ spread out to $\la = \infty$. In agreement with the phase
diagram drawn by Roan and McCoy \cite{roan}, we conclude that in the
integrable case, phase (I) extends up to $\vph=\pi$ and $\la \approx 1.6$.
\end{itemize}  \par For the self-dual case we are able to determine
the boundary between phases (I) and (IV) with good precision,
because the lowest gaps in region (I) are exponentially small as mentioned
above, whereas the gaps in the IC-region are generally of order unity.
In Tables~1 and~2 we give an idea of the precision with which the phase
boundary to the
IC-phases can be determined by our finite-size extrapolation \cite{GK}. \\
\zeile{1}  \begin{tabular}{|r|ccccccc|} \hline
\multicolumn{1}{|c|}{ } & \multicolumn{7}{c|}{$\phi=\vph$ (degrees)} \\
 \hline  $N$ & $\lambda=$0.1 & 0.25 & 0.5 & 0.6 & 0.7 & 0.8 & 0.9 \\   \hline
9 &170.860&157.553&135.605&127.301&119.986&114.519&112.333\\
10&171.112&158.085&135.767&126.779&118.457&111.796&108.522\\
11&171.358&158.649&136.252&126.751&117.556&109.763&105.366\\
12&171.589&159.208&136.927&127.060&117.121&108.263&102.728\\
\hline
$\infty$&170.43(4)&157.1(1)&135.7(2)&126.(3)&115.1(1)&103.7(1)&87.6(7)\\
\hline       \end{tabular}      \zeile{1}
Table~1. : Determination of the phase boundary between the high-temperature
phase $II$ and the IC-phase $IV$ from the first crossing of the levels
$0_0$ and $1_1$. For brevity, we quote only the values for $N=9,\ldots,12$.
However, in estimating the limiting values denoted "$\infty$",
data for all sites $N=3,\ldots,12$ have been used. The estimated errors
of the last given digit are quoted in brackets.
\\  \zeile{1}  \begin{tabular}{|r|ccccccc|} \hline
\multicolumn{1}{|c|}{ } & \multicolumn{7}{c|}{$\phi=\vph$ (degrees)} \\
 \hline  $N$ & $\lambda=$1.1 & 1.2 & 1.4 & 1.8 & 2.2 & 2.6 & 3.0 \\
\hline
9 &120.007&124.409 &132.599 &143.899 &150.729 &155.270 &158.526\\
10&116.541&121.733 &131.092 &143.357 &150.491 &155.149 &158.457\\
11&113.703&119.695 &130.146 &143.192 &150.521 &155.232 &158.553\\
12&111.355&118.152 &129.613 &143.278 &150.718 &155.438 &158.746\\
\hline
$\infty$& 99.(1)&113.2(1)&127.8(2)&144.(1)&151.(2)&156.(2)&159.(1) \\
\hline       \end{tabular}    \zeile{1}
Table~2. : Determination of the phase boundary between the low-temperature
phase $I$ and the IC-phase $III$ from the first crossing of the levels
$1_0$ and $2_2$.
\zeile{1}
Whether these transition lines have the properties of a
Pokrovski-Talapov type transition
\cite{pok} is an interesting open question, because here {\em two}
modulated phases have to coexist at the phase boundary line.

\section{Determination of the wave vector scaling exponent}
In this section we shall take a closer look at the $N$-dependence of the level
crossings in the low-temperature region $\la>1$.
In Table ~3 we give the cross-over positions in $\la$ for the
superintegrable case which appear for $N$ up to $12$ sites.
\par This Table shows that the cross-over
lines converge towards $\la=1$ and not towards the I/III-phase boundary
$\la=1.109518$. So the low-lying $p=0$-levels and their cross-overs do not
care about the point where the IC-phase sets in.
Between $\la=0.901293$ and $\la=1/0.901293$ just the $p=0$-levels are no
longer the lowest levels. This is analogous to the formula (\ref{lin}) for
the $p=0$-gap valid for $\la<1$ which also takes no notice of the
point $\la=0.901293$.
\zeile{1} \begin{tabular}{|c|rrrrr|} \hline
 $N$ &$Q\!=$0/1/2\hspace*{11mm} &\hspace*{-7mm} 1/0/2 \hspace*{14mm} &
                                 \hspace*{-9mm} 1/2/0 \hspace*{14mm} &
 \hspace*{-9mm} 2/1/0 \hspace*{14mm} & \hspace*{-9mm} 2/0/1 \hspace*{14mm}
 \\  \hline
 3   & 1.8789156 &           &           &            &        \\
 4   & 1.3645610 &           &           &            &        \\
 5   & 1.2081036 & 3.3430364 &           &            &        \\
 6   & 1.1368767 & 2.0826986 &           &            &        \\
 7   & 1.0978185 & 1.6742233 & 4.8041526 &            &        \\
 8   & 1.0738838 & 1.4761904 & 2.8030552 &            &        \\
 9   & 1.0580721 & 1.3610533 & 2.1459063 & 6.2689052  &        \\
10   & 1.0470391 & 1.2866459 & 1.8229750 & 3.5290973  &        \\
11   & 1.0390113 & 1.2350807 & 1.6327694 & 2.6244231  &        \\
12   & 1.0329732 & 1.1975179 & 1.5083542 & 2.1772063  & \hspace*{-11mm}4.255
\\
\hline  $\infty$& 0.9997(5) & 0.999(1) & 0.990(6) & 0.93(5)   &  \\  \hline
\end{tabular}   \zeile{1}
Table 3. : Positions in $\la$ of the $p\!=\!0$-sector lowest level
crossings on the superintegrable line $\vph=90^\circ$ for periodic b.c.  In the
top line the order of the levels with different charges $Q$
in the $\la$-interval between the cross-overs is given. So the
first column of $\la$-values gives the position of the $0_0/1_0$-cross-over,
the
second column the $0_0/2_0$-cross-over in the first excited state, etc.
\zeile{1}
Critical exponents of the translationally invariant part of the
spectrum at \mbox{$\la=1$} have been discussed in \cite{al5,ba1}.
These authors find that scaling is asymmetrical around $\la=1$ and derived the
specific heat exponent to be $\alpha=1-2/n$ for the $Z_n$-superintegrable
models. From (\ref{lin}) it is immediate that the magnetic critical index
is $\nu=1$. The spacial correlation exponent $\nu_s$ has been calculated
by Albertini {\em et al.} \cite{al5} from the $\la$- and $N$-dependence of
the $Q\!=\!0,p\!=\!0$-"ground"-state alone to be $\nu_s=2/n$. Baxter
\cite{ba1} finds from the exact solution of the superintegrable model the same
value $2/n$ for the interfacial tension exponent which desribes the
exponential approach in $N$ to degeneracy of the $p=0$-lowest levels
in the low-temperature phase near $\la=1$.
\par We want to show that in the $Z_3$-case (\ref{H3}) one obtains
$\nu_s=\frac{2}{3}$ for the {\em wave vector exponent} at $\la=1$. We are not
restricted by our method to the superintegrable point and find that $\nu_s$
varies little or not at all for a wide range of the chiral angle $\vph$.
All our results are consistent with the hypothesis that
it is the inverse wave vector which sets the dominant spacial scale length
of the $p=0$ sector of the spectrum around $\la=1$.
\par The finite-size scaling rules for quantum chains with oscillating phases
have been formulated in \cite{hoeg}. Let us define $\nu_s$ as the power of
divergence of the inverse wave vector as $\la \ra \la_c$:
\BEQ   1/K \sim (\la-\la_c)^{-\nu_s}.  \EEQ
In the region dominated by the length scale $1/K$ we use the scaling
variable   \BEQ z = N^{1/\nu_s}(\la -\la_c)   \EEQ
{}From Table~3 we deduce that $\la_c =1$. In the scaling regime we can express
the cross-over positions $\la_k(N)$ ($k=1,2,\ldots$) in terms of values $z_k$
of
the scaling variable. Calling $\Lambda_{k,k'}(N)$ the distance between two
level
crossing points (e.g. at fixed $\vph$) we can write
\BEA
\Lambda_{k,k'}(N) & \equiv & \la_k(N) - \la_{k'}(N) \nonumber
        \\ & = &   (z_k- z_{k'}) N^{-1/\nu_s}.    \label{lk}   \EEA
Using eq.(\ref{lk}) for $N$ and $N-1$ sites, we get $N$th approximants
for $\nu_s$:
\BEQ     \nu_s(N,k,k') = \frac{\ln(N/(N-1))} {\ln{\frac{\Lambda_{k,k'}(N-1)}
 {\Lambda_{k,k'}(N)}}}.  \label{nuf}   \EEQ
If our assumptions are consistent, the large $N$-limit of these
approximants has to give the same value $\nu_s$ for every choice of pairs
$k,k'$:
\BEQ   \nu_s = \lim_{N \ra \infty} \nu_s(N,k,k'). \EEQ
One may also refer the distances to $\la_c=1$ and use
$\Lambda_{k,0}(N) = \la_k(N) -\la_c$ in (\ref{nuf}), corresponding to $z_0=0$
in
(\ref{lk}).
\par In Table~4 we give numerical results from three different choices
of $\Lambda_{k,k'}$.
Observe that from $\vph=120^\circ$ down to $\vph=60^\circ$ we find no
appreciable variation of $\nu_s$ with $\vph$. In particular, there is no
indication of a rise towards the $\vph=0$-Potts value $\nu_s=\frac{5}{6}$.
Unfortunately, for small $\vph$ the cross-over lines move to large $\la$ for
our small $N$ available, so that the convergence of the $\nu_s(N,k,k')$ becomes
poor e.g. for $\vph=30^\circ$.
\par From the level crossing points and the slopes of the gaps at these points,
we can also obtain information on the specific heat exponent $\alpha$ and the
magnetic gap exponent $\nu$ \cite{hoeg}. The specific heat $C_v$ and its
exponent $\alpha$ are defined by
\BEQ C_v(\la) = -\frac{1}{N}\frac{d^2 E_0}{d\la^2} \sim |\la_c -\la|^{-\alpha}.
         \label{Cv}  \EEQ                     In the region with
ground-state cross-overs, its finite-size approximants $C_v(\la,N)$ are
\newpage
\begin{tabular}{|r|ccc|cccc|c|} \hline
\multicolumn{1}{|c|}{ } & \multicolumn{3}{c|}{$\Lambda_{1,0}=\la_{0/1}-\la_c$ }
  & \multicolumn{4}{c|}{$\Lambda_{2,0}=\la_{0/2}-\la_c$} & \multicolumn{1}{c|}
{\hspace*{-3mm} $\la_{0/1}-\la_{0/2}$\hspace*{-1mm}}  \\
\hline $N$&$\vph=60^\circ$&$75^\circ$&$90^\circ$&$60^\circ$&$90^\circ$&
$105^\circ$&$120^\circ$ & $90^\circ$ \\  \hline
 5& 0.2870& 0.3741& 0.3980&       &        & 0.1997& 0.2809&       \\
 6& 0.3800& 0.4323& 0.4352&       &  0.2362& 0.3156& 0.3837& 0.2239 \\
 7& 0.4352& 0.4668& 0.4588&       &  0.3255& 0.3863& 0.4561& 0.3113 \\
 8& 0.4712& 0.4893& 0.4758& 0.1809&  0.3840& 0.4354& 0.5096& 0.3713 \\
 9& 0.4961& 0.5050& 0.4891& 0.2636&  0.4255& 0.4719& 0.5489& 0.4154 \\
10& 0.5141& 0.5165& 0.5000& 0.3232&  0.4565& 0.5000& 0.5773& 0.4490 \\
11& 0.5276& 0.5252& 0.5093& 0.3682&  0.4806& 0.5223& 0.5977& 0.4753 \\
12& 0.5380& 0.5321& 0.5174& 0.4032&  0.4998& 0.5403& 0.6125& 0.4964 \\
\hline
$\infty$&0.59(1)&0.59(2)& 0.66(4)& 0.61(2)& 0.67(4)& 0.66(2)&0.66(1)&0.66(1)\\
\hline       \end{tabular}
\zeile{1}     Table 4. : Approximants $\nu_{k,k'}(N)$ to
the wave-number exponent $\nu_s$ using three different levels choices of
$\Lambda_{k,k'}$: First $\Lambda_{1,0}$, which is the distance of the first
ground-state cross-over from $\la=1$, then $\Lambda_{2,0}$, the distance of
the first cross-over in the first excited state from $\la=1$ and, in the last
column, the distance $\Lambda_{2,1}$ between the first cross-overs in the
ground state and first excited state. \zeile{2}
\noindent   given
by a sum of $\delta$-function contributions (neglecting some smooth
background):
\BEQ  C_v(\la,N) = \sum_{k=1}^{k_{max}(N)} C_k(N)\delta(\la-\la_k(N))
     \label{cdel}    \EEQ
The weight factors of the $\delta$-terms are given by the $\la$-derivatives
of the magnetic gaps at the ground state cross-over positions:
\BEQ C_k(N)=\frac{1}{N}\left|\frac{d(E_{Q_i}-E_{Q_j})}{d\la}\right|_{\la_k(N)}.
     \label{cdl}  \EEQ   The relevant charge sectors $Q_i, Q_j$ rotate as
$\la$ increases, see Table~3. \par
In the scaling regime we express $C_v(\la,N)$ in terms of the specific heat
scaling function ${\cal C}(z)$      \BEQ
C_v(\la,N) = N^{\alpha/\nu_s} {\cal C}(z) = N^{\alpha/\nu_s} \sum \tilde{C}_k
   \delta(z - z_k),     \EEQ
so that \BEQ    C_k(N) = \tilde{C}_k N^{(\alpha-1)/\nu_s}. \label{ccx} \EEQ
{}From (\ref{ccx}) we get approximants $\alpha(N,k)$ for $\alpha$:
\BEQ  \alpha(N,k) -1 = \nu_s(N,k)\ln\left(\frac{C_k(N-1)}{C_k(N)}\right)/
 \ln\left(\frac{N-1}{N}\right)   \label{alk}   \EEQ
Formula (\ref{alk}) gives a practicable method for calculating $\alpha$ from
the $\nu_s(N,k)$ and the ($\Delta Q \neq 0$)-magnetic gap slopes (\ref{cdl}).
\par
The fact that the specific heat contributions from the ground-state crossings
are proportional to magnetic gap slopes gives rise to the generalized
hyperscaling relation \cite{hoeg}-\cite{dom}:
  \BEQ \alpha+\nu_s+\nu=2, \label{hy} \EEQ
which is easily seen to be valid even for the finite-$N$-approximants to
$\alpha, \nu_s$ and $\nu$ \cite{hoeg}.
\par  For detailed tables of the relevant sequences of approximants, we refer
the reader to \cite{GK}. Here we just mention that we find $\nu=1$ within
a few percent error for all $60^\circ \le \vph \le 120^\circ$. So
for these values of $\vph$ eq.(\ref{hy}) reads $\frac{1}{3} + 1 + \frac{2}{3}
= 2$. For the Potts case $\vph = 0^\circ$ we know that it must read
$\frac{1}{3}
+\frac{5}{6} + \frac{5}{6} = 2$, but we are not able to tell at which $\vph$
the
first equation goes over to the second one.

\zeile{1}
\section{Particle interpretation of the low-lying spectrum}
In this section we attempt to describe the low-lying part of the spectrum of
$\HH$ of eq.(\ref{H3}) in terms of elementary excitations. For the
superintegrable case such a discussion has been given e.g. in \cite{Mc91,das},
but it is instructive to consider the dependence on the chiral angle too.
We shall treat the self-dual case $\vph=\phi$ if not stated otherwise.
\par Let us start with the information we have for $\vph=0$. This is the
standard $Z_3$-Potts model which has a second order phase transition at $\la=1$
decribed by a $c=\frac{4}{5}$-conformal field theory \cite{dot,grp}.
Zamolodchikov has shown \cite{zaz3} that the high-temperature scaling regime
around $\la=1$ is described by the thermal perturbation of the $c=\frac{4}{5}$-
conformal
theory, which, as he showed, is still integrable. The $S$-matrix of this
massive
theory factorizes and contains just a single pair of particles with
$Z_3$-charges $Q\!=\!1$ and $Q\!=\!2$. Because of the additional $S_3$-symmetry
present at $\vph=0$ the $Q\!=\!1$ and $Q\!=\!2$-sectors are degenerate and
consequently both particles have the same mass, forming a
particle-antiparticle pair. The higher levels of the high-temperature scaling
Potts-model are multiparticle scattering states of which even the phase shifts
are known.
\par In order to find out whether the spectrum for $\vph \neq 0$ is
continuously
obtained from the Potts spectrum, we first take a look at a high-temperature
expansion of the $p=0$-states. In the interval $-\frac{\pi}{2}<\vph<\pi$ we
find
for the lowest $Q=1$,~$p=0$-gap up to order $\la^2$:
\BEQ  \Delta E_{1,0}(\vph,\la) = 4 \sin\frac{\pi -\vph}{3} -
 \frac{4\la}{\sqrt{3}}\cos\frac{\vph}{3}
-\frac{2\la^2}{\sqrt{3}}\:\frac{\sin{{\textstyle \frac{1}{3}}(\pi -2\vph)}}
{\sin{{\textstyle \frac{1}{3}}(\pi +2\vph)}}\, f(\vph) +\ldots \label{Q1} \EEQ
Here $f(\vph)$ is a smoothly varying function of $\vph$, which satisfies
\BEQ  f(-{\textstyle \pih}) = 2/\sqrt{3};~~~f(0) = 1;~~~f({\textstyle \pih})
 = \sqrt{3}/2;~~~
\lim_{\vph\ra\pi} f(\vph)/\sin{{\textstyle \frac{1}{3}}(\pi-\vph)}=2\sqrt{3}.
 \label{fi}        \EEQ   Applying a $CP$-transformation, the same formula can
be used for the $Q\!=\!2$, $p=0$-gap if $|\vph| < \pih$:
\BEQ  \Delta E_{2,0}(\vph,\la)=\Delta E_{1,0}(-\vph,\la). \label{Q2}   \EEQ
For small values of $\la$, for which these formulae are useful, the
level \mbox{
$E_0(Q\!=\!1,p\!=\!0)$} is isolated in the spectrum of the $Q\!=\!1$-charge
sector if $0 \leq \vph < \pi$. Similarly,the level
$E_0(Q\!=\!2,p\!=\!0)$ is isolated in the spectrum of the charge sector
$Q\!=\!2$ for $0\le\vph < \pih$. So we tentatively call these two
levels "single particle levels" and assign the masses       \BEQ
m_1\equiv\Delta E_{1,0} ; \hspace{10mm} m_2\equiv\Delta E_{2,0}. \label{m} \EEQ
For $\vph \ra 0$ both particles become degenerate, in agreement with our
discussion of the Potts case.
\par  Now let us take a look at formulae (\ref{Q1})~(\ref{Q2}) for $\vph\ra
\frac{\pi}{2}$. Disregarding first the $\la^2$-term, with
$\vph$ increasing, $m_2$ increases faster than $m_1$ decreases. For $\vph=
\frac{\pi}{2}$ (the superintegrable case) we seem to get
\BEQ   m_1=2-2\la +O(\la^2);\hspace{18mm} m_2=4-2\la +O(\la^2).\EEQ
While this formula is allright for $m_1$ (we even know that all higher order
terms in $\la$ vanish as $\vph\ra\pih$) it is not correct for $m_2$, since
for $m_2$ the coefficient of $\la^2$ diverges as $\vph\ra\pih$.
Physically this divergence is not surprising because as $\vph\ra\pih$
particle $m_2$ moves towards the threshold of the
two-$m_1$-particle scattering states which are in the same $Q\!=\!2$-sector,
since the $Z_3$ charges of the particles are additive {\em mod} 3.
\par Calculating numerically $m_1$ and $m_2$ for many points of the
region \mbox{$0\le\vph < \pih$},~~
$0\le\la < 1$ we find that there is a smooth change of the ratio $m_2/m_1$
from its Potts-value $m_2/m_1 = 1$ to
$m_2/m_1 \ra 2$ as $\vph$ approaches $\pi/2$ from below.
\par Since we can easily calculate numerically the about eight lowest $p=0$
levels of each charge sector for all $0\le\vph\le\pi$, we now check whether
the pattern of higher levels
fits to the proposed particle interpretation. In Figs.~3 and ~4
we show the expected patterns of levels and thresholds,
which should be qualitatively different for $\vph<\pih$ and
$\vph\ge\pi$ if in the latter case $m_2$ becomes larger than $2m_1$.
\begin{itemize} \item Sector $Q=0$: As long as $m_2 < 2m_1$ ($\vph<\pih$), the
first threshold is due
to the scattering of two different particles $m_1$ and $m_2$, followed
at $\Delta E=3m_1$ by the 3-$m_1$-particle scattering states (here again
we refer the energy gaps to the $Q\!=\!0,~p\!=\!0$-ground state).
This two-particle threshold moves into the $3m_1$-continuum at the
superintegrable line if there we have $m_2 = 2m_1$.
In Table~5 we give the finite-$N$ energy gaps accessible to our numerical
diagonalization of $\HH$ for the superintegrable case and $\la=0.5$.
We choose this value of $\la$ because then from (\ref{lin}) we have $m_1=1$ and
a normalization
of the data is unnecessary. The superintegrable case is particularly
interesting, because it makes it necessary to disentangle physically different
but degenerate levels.   \par Not surprisingly after this general discussion,
in our numerical calculation, the lowest bunch of levels appears at $\Delta
E=3m_1$. Now, in order to find out, which of these levels are $m_1+m_2$-states
and which are the $m_1+m_1+m_1$-states, we look into their finite-size
behaviour. For a single particle level we expect exponential convergence
for $N\ra\infty$:
\BEQ \Delta E(N)-\Delta E(\infty) = \exp(-N/\xi) +\ldots    \label{exx} \EEQ
where $\xi$ is a correlation length. In contrast, for multiparticle
scattering states we expect power convergence \cite{LuW,sag}:
\BEQ  \Delta E(N)-\Delta E(\infty) \sim N^{-y} + \ldots  \label{mul} \EEQ
We calculate approximants $y_n$ to $y$ using neighbouring values of $N$:
\BEQ y_N = \ln\left(\frac{\Delta E(N)-\Delta E(\infty)}{\Delta E(N-1)-
   \Delta E(\infty)}\right)/\ln\left(\frac{N}{N-1}\right). \label{yn} \EEQ
Table~6 shows some $y_n$ calculated from the $N$-dependence of the energy
levels in Tab.~5. We see that the limiting values of $y$ come out clearly to
be very close either to $y=2$
or $y=3$ and allow us to decide which of the levels are $m_1+m_2$-states
and which are $m_1+m_1+m_1$.
\item Sector $Q=1$: Here we see exponential convergence for the isolated
level $m_1$, and with good precision we see at $\Delta E=4$ the bunch of the
next higher levels. There are levels converging with $y=2$ and others
converging with $y=3$. The obvious interpretation is in terms of $m_2+m_2$-
two-particle states and of $m_1+m_1+m_2$-three particle states. We have checked
that moving towards smaller values of $\vph$ these levels split in the expected
way: for $m_2<2m_1$ the levels interpreted as "two-particle" go lower than the
assumed three-particle states.
\item Sector $Q=2$: For e.g. $\vph=45^\circ$ there is the clear
exponential convergence of the isolated $m_2$-level. However, the interesting
and crucial test comes now in the superintegrable case:
Among the several levels at
$\Delta E=2$ (we still consider $\la=0.5$) which show power convergence, can we
distinguish one level which shows exponential behaviour? Already without
calculating the sequences $y_n$ we see that the level in the first row
of the bottom part of Table~5 sticks out by its very fast convergence.
This is substantiated in Table~6 where we see the corresponding sequence of
the $y_n$ growing very fast.
So we have evidence that the particle $m_2$ survives up to $\vph=\pih$.
We have tried to follow $m_2$ into the continuum for $\vph>\pih$. For
$\vph=105^\circ$ and $\la=0.15$ we have succeeded in distingishing
one level which converges faster than a low power $y=2$ or $y=3$.
\par We have also performed calculations for $\vph=45^\circ,~105^\circ$, both
for the self-dual and for the integrable model. In all these cases we find
a pattern of the ca.6 lowest levels of each charge sector as scetched in
Figs.~3 and 4.
\end{itemize}  \par By analogy, we expect that the superintegrable $Z_n$-models
have low-lying excitations corresponding to $n-1$ particle species, one in each
$Q \neq 0$-sector with mass $m_Q = Q m_1$. \zeile{1}
Fig.~3: ({\bf following page}) General structure of the low-lying spectrum.
Left
hand side of the picture: Structure of the spectrum in the different charge
sectors if the chiral angle $\vph$ has a value below $\vph=\pih$.
Right hand side: for $\vph$ above the superintegrable value.
\newpage   \vspace*{5cm}  \hspace*{4cm}   (space for Fig.~3)  \newpage
\begin{tabular}{|r|cccccc|} \hline
\multicolumn{1}{|c|}{ } & \multicolumn{6}{c|}{$\Delta E_{Q=0,i}$} \\
        $N$& $i=1$ & 2 & 3 & 4 & 5 & 6 \\  \hline
 6& 3.2402540& 5.4035327& 4.7691472& 6.4717590& 6.2844859&          \\
 7& 3.1598983& 4.8359585& 4.4520381& 5.9303676& 5.9032240& 7.0487030\\
 8& 3.1115649& 4.4240923& 4.2075367& 5.4868664& 5.5496714& 6.3987521\\
 9& 3.0808417& 4.1213280& 4.0168334& 5.1252880& 5.2384091& 5.8572524\\
10& 3.0604137& 3.8954662& 3.8661605& 4.8297326& 4.9700825& 5.4121174\\
11& 3.0463166& 3.7244809& 3.7455652& 4.5867795& 4.7406004& 5.0471104\\
12& 3.0362806& 3.5932058& 3.6478379& 4.3856960& 4.5446740& 4.7471645\\
\hline $\infty$&
 2.9998(3)& 2.986(7) & 3.01(1)  & 3.02(8)  & 3.05(6)  & 2.9(1)   \\
\hline  $y$ &3.0(1)&2.9(2)&2.0(1)&2.1(1)&1.9(1)&2.8(6)     \\  \hline
\end{tabular}
\zeile{1}
\begin{tabular}{|c|ccccc|} \hline
\multicolumn{1}{|c|}{ } & \multicolumn{5}{c|}{$\Delta E_{Q=1,i}$} \\
        $N$& $i=0$ & 1 & 2 & 3 & 4 \\  \hline
 6& 0.9963767& 4.7353052& 6.1998192& 7.2942778& 6.5428730\\
 7& 0.9984874& 4.5173085& 5.7905491& 6.6049711& 6.1612566\\
 8& 0.9994289& 4.3760023& 5.4775432& 6.0785242& 5.8456347\\
 9& 0.9998044& 4.2809972& 5.2349161& 5.6757753& 5.5870148\\
10& 0.9999405& 4.2150750& 5.0442562& 5.3652765& 5.3749672\\
11& 0.9999851& 4.1680556& 4.8924532& 5.1235743& 5.2002368\\
12& 0.9999978& 4.1336962& 4.7700872& 4.9335093& 5.0552644\\
\hline $\infty$&
 1.0000000& 3.99(2)  & 4.00(1)  & 4.01(5)  & 4.01(3)     \\
\hline  $y$ &expon.&3.0(2)&2.0(1)&2.9(2)&1.9(1) \\  \hline
\end{tabular}
\zeile{1}
\begin{tabular}{|r|cccccc|} \hline
\multicolumn{1}{|c|}{ } & \multicolumn{6}{c|}{$\Delta E_{Q=2,i}$} \\
        $N$& $i=0$ & 1 & 2 & 3 & 4 & 5 \\  \hline
 6& 2.0002000& 2.9975868& 4.8074676& 5.9788475&          & 6.5436728\\
 7& 1.9995726& 2.7901373& 4.3798245& 5.6633417&          & 6.1294869\\
 8& 1.9996761& 2.6389905& 4.0256344& 5.3106885& 6.0425834& 5.8468390\\
 9& 1.9998304& 2.5261089& 3.7351818& 4.9697753& 5.8323103& 5.6487379\\
10& 1.9999247& 2.4399607& 3.4970152& 4.6583108& 5.5766919& 5.5065400\\
11& 1.9999701& 2.3729391& 3.3008958& 4.3808694& 5.3110577& 5.4022604\\
12& 1.9999893& 2.3198948& 3.1384071& 4.1365977& 5.0527923& 5.3242970\\
\hline $\infty$& 2.0001(1)& 2.002(7) & 2.00(5) & 2.0(1) & 2.6(5) & 4.99(1) \\
\hline $y$ &expon.&2.0(1)&1.9(1)&1.9(2)&  ?   &2.9(2)\\  \hline
\end{tabular}
\zeile{1}
Table~5: The lowest energy gaps $\Delta E_{Q,i}$, as defined in eq.(\ref{gap}),
for the $Z_3$-ha\-mil\-tonian eq.(\ref{H3}), for the superintegrable case
$\phi = \vph = \pi/2$, and $\lambda = 0.50$.
Upper Table: sector $Q=0$, middle table: sector $Q=1$ and bottom table: sector
$Q=2$.
The numbers given in brackets indicate the estimated error in the last
written digit.
\newpage
\begin{tabular}{|c|ccc|ccc|ccc|} \hline  \multicolumn{1}{|c|}{ } &
        \multicolumn{3}{c|}{$y_N$ for $\Delta E_{Q=0,i}$} &
        \multicolumn{3}{c|}{$y_N$ for $\Delta E_{Q=1,i}$} &
        \multicolumn{3}{c|}{$y_N$ for $\Delta E_{Q=2,i}$} \\
   $N$ & $i=1$ & 2 & 3  &$i=0$ & 1 & 2 &$i=1$& 2 & 3\\  \hline
  5& 2.443 &        & 0.984 &  2.092  & 1.956 & 1.040 &        &
                                                            1.284 & 0.654 \\
  6& 2.563 & 1.552 & 1.153 &   4.046  & 2.141 & 1.206 &        &
                                                           1.413 & 0.897 \\
  7& 2.641 & 1.748 & 1.281 &   5.667  & 2.181 & 1.335 &        &
                                                            1.512 & 1.072 \\
  8& 2.696 & 1.902 & 1.381 &   7.295  & 2.389 & 1.439 & 2.078 &
                                                            1.590 & 1.207 \\
  9& 2.735 & 2.029 & 1.459 &   9.095  & 2.473 & 1.523 & 5.492 &
                                                            1.650 & 1.314 \\
 10& 2.765 & 2.135 & 1.522 &   11.30 & 2.538 & 1.592 & 7.703 &
                                                            1.697 & 1.401 \\
 11& 2.788 & 2.223 & 1.573 &   14.51 & 2.588 & 1.648 & 9.699 &
                                                            1.734 & 1.473 \\
 12& 2.807 & 2.298 & 1.615 &   21.81 & 2.629 & 1.695 & 11.79&
                                                            1.763 & 1.533 \\
\hline  $\infty$ &3.0(1)&
 2.9(2)&2.0(1)&expon.&3.0(2)&2.0(1)&expon.&2.0(1)&1.9(1)\\  \hline
\end{tabular}
\zeile{1}
Table~6: Exponents $y_N$ of the convergence $N \rightarrow \infty$ as defined
in eq.(\ref{yn}), for the superintegrable case and $\lambda =0.50$ for the
lower
levels given in Table 5. "expon." in the bottom row means that the increasing
sequence of the $y_N$ in the column above indicates exponential convergence.
\zeile{1}
\par Having been successful with the two-particle interpretation of the
low-lying spectrum, we finally consider the energy-momentum relation of these
particles. Since parity is generally not a good quantum number of our model,
the dispersion relation is not expected to be symmetrical between positive
and negative momenta. In \cite{al2} such unsymmetrical dispersion curves have
been given for the superintegrable case. We have studied the $\vph$-dependence
of this asymmetry. Of course, for $\vph\ra 0$ the asymmetry must vanish
because for $\vph=\phi=0$ the hamiltonian $\HH$ is parity invariant. The
high-temperature expansion gives us some information about the increase
of the asymmetry with $\vph$ (we consider again the self-dual case):
\par  For calculating the limit $N \ra \infty$ using lattice momenta $p$ as in
eq.(\ref{p}), we introduce the macroscopic momentum $P$ defined by
\BEQ    P = \frac{2\pi}{N}\,p \hspace{15mm} (-\pi \geq P \geq \pi)      \EEQ
Then, in generalisation of (\ref{Q1}),~(\ref{Q2}) we find up to order $\la$:
\BEQ  \Delta E_{1,0}(\lambda,P) = 4 \sin\frac{\pi -\vph}{3}
 -\frac{4\lambda}{\sqrt{3}} \cos (P-{\textstyle \frac{1}{3}}\vph)
            + \ldots~~~~~(0\leq\vph<\pi)                   \label{P1} \EEQ
and
\BEQ  \Delta E_{2,0}(\lambda,P) = 4 \sin\frac{\pi +\vph}{3}
 -\frac{4\lambda}{\sqrt{3}} \cos (P+{\textstyle \frac{1}{3}}\vph)
           + \ldots~~~~~(0\leq\vph<\pi/2).                \label{P2} \EEQ
This shows that for small $\la$, the asymmetry of the energy-momentum relation
with respect to $P \ra -P$ is due only to the presence of a macroscopic
momentum
$P_m=\pm\vph/3$ (observe that this is the same momentum which appears
in eq.(\ref{mac})
for $\vph=\pih$). We have checked numerically that for $\la\le 0.1$ these
formulae are an excellent description of the data.
Only for large values of $\la$ the situation is getting
complicated. \par Of course, since we considered just a few low-lying levels,
we
are not able to make statements about higher levels. In ref.\cite{das}
it has been shown that in the superintegrable case
the model contains levels which cannot be interpreted as
quasiparticles in the sense that their energies and momenta receive unrelated
additive contributions.
\section{Conclusions}   In this talk we reported results of a numerical
investigation of the low-lying spectrum of the
$Z_3$-chiral Potts hamiltonian (\ref{H3}) for $N=2,\ldots,12$ sites.
We get an overview about the dependence of ca. 18 levels
on the chiral angle and the inverse temperature. This fits to the
interpretation of the high-temperature low-lying spectrum
in terms of two basic particles with $Z_3$-charges
$Q\!=\!1$ and $Q\!=\!2$ and their scattering states.
The superintegrable case is distinguished by the special 1~:~2-mass  ratio of
the particles. Even when moving slightly into the continuum formed by two
$Q\!=\!1$-particles, the heavy $Q\!=\!2$-particle
can be distinguished by its exponential finite-size behaviour.
Two- and three-particle scattering states can be distinguished by their
different finite-size power behaviour.
High-temperature perturbation theory gives a simple expression for the
effect of parity violation on the energy-momentum relation of the particles.
\par Particular attention has been devoted to the
study of ground-state level crossings in the low-temperature regime. We
found that the massive phases have
oscillatory correlation functions and calculated the wave-vector critical
exponent from the low-temperature phase. It turns out that for translation
invariant features of the model in the neighbourhood of the self-dual line,
the diverging inverse wave vector defines
the most relevant length scale.

\zeile{1}  \noindent {\large {\bf Acknowledgements:}} \hspace*{3mm}
The results reported in Sects.~2~and~3 of this talk have been obtained in
collaboration with Torsten Krallmann. I am grateful to T. Krallmann for
numerous valuable discussions. I wish to give my special thanks to Professor
\mbox{J. Q. Liang} for his kind hospitality in Taiyuan during the ISATQP
symposium. \zeile{2}

\end{document}